\documentclass{kapprocmod}
\usepackage{graphicx}
\usepackage{amsmath}
\let\footnote\savefootnote
\let\footnotetext\savefootnotetext
\let\footnoterule\savefootnoterule
\kluwerbib
\startauthorindex
\def\ome{\boldsymbol{\omega}}
\def\xib{\boldsymbol{\xi}}
\begin{document}
%
\articletitle[Numerical evidence of breaking of vortex lines in an ideal fluid]
{Numerical evidence of breaking\\ of vortex lines in an ideal fluid}
\booktitlerunninghead{E.~A.~Kuznetsov, O.~M.~Podvigina and V.~A.~Zheligovsky}
\chaptitlerunninghead{Numerical evidence of breaking of vortex lines}
\author{EVGENIY A.~KUZNETSOV\nobreak
$^{1,3}$,
OLGA M.~PODVIGINA$^{2,3}$\\
\& VLADISLAV A.~ZHELIGOVSKY$^{2,3}$
}\vskip6pt
%
%
\affil{\hspace*{15pt}
$^1$\nobreak			
L.D.Landau Institute for Theoretical Physics,\\
\hspace*{20pt}			
2 Kosygin str., 117334 Moscow, Russian Federation
\\ \hspace*{22pt}
{\emailfont kuznetso@itp.ac.ru}
}\vskip6pt
\affil{\hspace*{14pt}
$^2$
International Institute of Earthquake Prediction Theory\\
\hspace*{20pt}
and Mathematical Geophysics,\\
\hspace*{20pt}
79 bldg.~2 Warshavskoe ave., 113556 Moscow, Russian Federation;\\
\hspace*{20pt}
Laboratory of general aerodynamics, Institute of Mechanics,\\
\hspace*{20pt}
Lomonosov Moscow State University,\\
\hspace*{20pt}
1, Michurinsky ave., 119899 Moscow, Russian Federation
}\vskip6pt

\affil{\hspace*{14pt}
$^3$
Observatoire de la C\^ote d'Azur, CNRS UMR~6529,\\
\hspace*{20pt}
BP~4229, 06304 Nice Cedex 4, France
}

\inxx{collapse}
\inxx{ideal incompressible fluid}
\inxx{vortex lines}
\inxx{breaking}
\inxx{breaking of vortex lines}
\inxx{singularity}
\inxx{singularity formation}
\inxx{point singularity}
\inxx{developed turbulence}
\inxx{Kolmogorov spectrum}
\inxx{statistical equations of hydrodynamics}
\inxx{intermittency}
\inxx{correlation functions}
\inxx{coherent structures in turbulence}
\inxx{vorticity}
\inxx{vortex tubes}
\inxx{flows with a spatial symmetry}
\inxx{Crow instability}
\inxx{Euler equation}
\inxx{force-free Euler equation}
\inxx{Euler equation for vorticity}
\inxx{Jacobian}
\inxx{Cauchy invariant}
\inxx{frozenness of vorticity}
\inxx{mixed Lagrangian-Eulerian description}
\inxx{integrable hydrodynamics}
\inxx{Taylor expansion}
\inxx{eigenvalue}
\inxx{``soft" direction}
\inxx{``hard" direction}
\inxx{self-similarity}
\inxx{scale}
\inxx{self-similar variables}
\inxx{pancake structure}
\inxx{hold}
\inxx{catastrophe}
\inxx{catastrophe theory}
\inxx{weak collapse}
\inxx{super-weak collapse}
\inxx{enstrophy}
\inxx{energy dissipation}
\inxx{energy dissipation rate}
\inxx{energy conservation}
\inxx{kinetic energy}
\inxx{viscosity}
\inxx{energy}
\inxx{Sobolev space}
\inxx{norm}
\inxx{theorem}
\inxx{necessary and sufficient condition for collapse}
\inxx{integrals of motion}
\inxx{numerical integration}
\inxx{$\bf R$-grid}
\inxx{$\bf r$-grid}
\inxx{$\bf a$-grid}
\inxx{moving grid}
\inxx{regular grid}
\inxx{linear interpolation}
\inxx{anisotropy}
\inxx{spatial resolution}
\inxx{numerical algorithm}
\inxx{numerical stability}
\inxx{ABC flow}
\inxx{eigenfunction}
\inxx{curl}
\inxx{steady solution}
\inxx{force-free Euler equation}
\inxx{solenoidal field}
\inxx{Fourier harmonics}
\inxx{random-amplitude Fourier harmonics}
\inxx{exponentially decaying spectrum}
\inxx{spectrum}
\inxx{smoothing}
\vspace*{-8pt}\begin{abstract}
Emergence of singularity of vorticity at a single point, not related to any
symmetry of the initial distribution, has been demonstrated numerically
for the first time. Behavior of the maximum of vorticity near the point of
collapse closely follows the dependence $(t_0-t)^{-1}$, where $t_0$ is
the time of collapse. This agrees with the interpretation of collapse
in an ideal incompressible fluid as of the process of vortex lines breaking.
\end{abstract}
\vspace*{-6pt}
\section{Introduction}
The problem of collapse in hydrodynamics, i.e. of a process of singularity
formation in a finite time, is essential for understanding of the physical
nature of developed turbulence. Despite a progress in construction of
statistical theory of Kolmogorov spectra within both diagram and functional
approaches (see, e.g., \cite{monin}; \cite{lvov} and references therein), so far
the question whether the Kolmogorov spectrum is a solution to the statistical
equations of hydrodynamics remains open. Another important problem, as yet
unsolved, is the one of intermittency. In statistical sense intermittency can
be interpreted as a consequence of a strongly non-Gaussian distribution of
turbulent velocity, resulting in deviation of exponents for higher correlation
functions from their Kolmogorov values (\cite{frisch}). Non-Gaussian behavior
implies that odd correlation functions do not vanish; this indicates the
presence of strong correlations between velocity fluctuations, suggesting
existence of coherent structures in turbulence. Analysis of both numerical and
experimental data reveals (see \cite{frisch} and references therein) that in
the regime of fully developed turbulence distribution of vorticity is strongly
inhomogeneous in space -- it is concentrated in relatively small regions. What
is the reason of this? Can such a high concentration be explained by formation
of singularity of vorticity in a finite time? How can one derive from this
hypothesis the Kolmogorov spectrum? This question is not rhetoric: it is well
known that any singularity results in a power-law kind of spectrum in the
short-scale region. Thus, the problem of collapse is of ultimate importance in
hydrodynamics.

The most popular object in the studies of collapse in hydrodynamics
is a system of two anti-parallel vortex tubes, inside which vorticity is
continuously distributed (\cite{kerr}), or in a more general setup -- flows
with a higher spatial symmetry (\cite{P2}; \cite{P1}). It is well known, that two
anti-parallel vortex filaments undergo the so-called Crow instability
(\cite{crow}) leading to stretching of vortex filaments in the direction
normal to the plane of the initial distribution of vortices
and to reduction of their mutual distance. It was demonstrated
in numerical experiments (\cite{kerr}) that point singularities are formed
in cores of each vortex tubes at the nonlinear stage of this instability,
and $|\ome|$ near the point of collapse increases like $(t_0-t)^{-1}$,
$t_0$ being the time of collapse (see also \cite{GMG}).

\section{Basic equations}

In this paper we present results\footnote{Preliminary results were communicated
in \cite{ZKP}.} of a numerical experiment, which can be
interpreted as emergence of singularity of vorticity at a single point
in a three-dimensional ideal hydrodynamic system, where initial data lacks
any symmetry. The representation of the Euler equation for
vorticity $\ome({\bf r},t)$ in terms of vortex lines is employed,
which was introduced in \cite{KR}:
\begin{equation}
\label{1}
\ome({\bf r},t)=(\ome_0({\bf a})\cdot\nabla_{\bf a}){\bf R}({\bf a},t)/J.
\end{equation}
Here the mapping
\begin{equation}
\label{2}
{\bf r}={\bf R(a,}t)
\end{equation}
represents transition to a new curvilinear system of
coordinates associated with vortex lines, so that
${\bf b}=(\ome_0({\bf a})\cdot\nabla_{\bf a}){\bf R}({\bf a},t)$
is a tangent vector to a given vortex line,
$J=\det\parallel\partial{\bf R}/\partial{\bf a}\parallel$ is the Jacobian of
the mapping (\ref{2}). Dynamics of the vector ${\bf R(a,}t)$ satisfies
\begin{equation}
\label{3}
\partial_t{\bf R}=\hat\Pi\,{\bf v(R,}t),
\end{equation}
where ${\bf v(R,}t)$ is the flow velocity at a point $\bf r=R$ and
$\hat\Pi$ is the transverse projection to the vortex line at this point:
\begin{equation}
\label{pro}
\Pi_{\alpha\beta}=\delta_{\alpha\beta}-\xi_{\alpha}\xi_{\beta},\quad
\xib={\bf b}/|{\bf b}|.
\end{equation}
Equations (\ref{1})-(\ref{3}) are closed by the relation
between vorticity and velocity:
\begin{equation}
\label{4}
\ome({\bf r},t)=\nabla\times{\bf v}({\bf r},t);\quad\nabla\cdot{\bf v}=0.
\end{equation}
The system of equations (\ref{1})-(\ref{4}) can be regarded as a result of
partial integration of the Euler equation
\begin{equation}
\label{5}
\partial_t\ome=\nabla\times[{\bf v}\times\ome],\quad\nabla\cdot{\bf v}=0.
\end{equation}
A vector field $\ome_0({\bf a})$ incorporated in (\ref{1}),
$\nabla_{\bf a}\cdot\,\ome_0({\bf a})=0$, is the Cauchy invariant, manifesting
frozenness of vorticity into the fluid. If \hbox{${\bf R(a},0)=\bf a$,}
$\ome_0$ is the initial distribution of vorticity.

The Jacobian $J$ can take arbitrary values because the description
under consideration is a mixed, Lagrangian-Eulerian one (\cite{KR}; \cite{KR00}).
In particular, $J$ can vanish at some point, which by virtue of (\ref{1})
implies a singularity of vorticity. It was demonstrated by \cite{KR00} that
collapses of this type are possible in the three-dimensional integrable
hydrodynamics \hbox{(\cite{KR}),} where in the Euler equation (\ref{5})
a modified relation between vorticity and velocity (both generalized) is assumed:
\begin{equation}
\label{int}
{\bf v}=\nabla\times(\delta{\cal H}/\delta\ome),\quad{\cal H}=\int|\ome|d\bf r.
\end{equation}
Emergence of singularity of vorticity at a point, where $J=0$,
means that a vortex line touches at this point another vortex line. This is
the process of breaking of vortex lines. Being analogous to breaking in a gas
of dust particles (dynamics of a gas with a zero pressure), this process
is completely determined by the mapping (\ref{2}).

\pagebreak
\section{Breaking of vortex lines}

Let us assume now that collapse in the Euler hydrodynamics occurs
due to breaking of vortex lines. Denote by $\tilde t({\bf a})>0$ a solution
to the equation $J({\bf a},t)=0$, and let $t_0=\min_{\bf a}\tilde t({\bf a})$,
where the minimum is achieved at ${\bf a}={\bf a}_0$.
Near the point of the minimum $(t_0,\ {\bf a}_0)$
the Jacobian can be expanded (cf.~\cite{KR00}):
\begin{equation}
\label{7}
J=\alpha(t_0-t)+\gamma_{ij}\Delta a_i\Delta a_j +...,
\end{equation}
where $\alpha>0$, $\gamma$ is a positive definite matrix and
$\Delta{\bf a}={\bf a}-{\bf a}_0$. The Taylor expansion (\ref{7}) is obtained
under the assumption that $J$ is smooth, which is conceivable up to the moment
of singularity formation. At $t=t_0$ the numerator in (\ref{1}),
i.e.~the vector ${\bf b}$, does not vanish: the condition $J=0$
is satisfied when the three vectors $\partial{\bf R}/\partial a_i$ ($i=1,2,3$)
lie in a plane, but generically none of them equals zero (that were
a degeneracy) so that near the point of singularity
\begin{equation}
\label{ome}
\ome({\bf r},t)\approx\frac{{\bf b}(t_0,{\bf a}_0)}
{\alpha(t_0-t)+\gamma_{ij}\Delta a_i\Delta a_j}.
\end{equation}
Furthermore, $J=0$ implies that an eigenvalue of the Jacoby matrix
(say, $\lambda_1$) vanishes, and generically the other two eigenvalues
($\lambda_2$ and $\lambda_3$) are non-zero. Therefore, there exist one ``soft"
direction associated with $\lambda_1$, and two ``hard" directions associated
with $\lambda_2$ and $\lambda_3$. It follows from (\ref{7}), that in the
auxiliary $\bf a$-space the self-similarity $\Delta{\bf a}\sim(t_0-t)^{1/2}$ is
uniform in all directions. However, in the physical space the scales are
different. Following \cite{KR00}, we show how an anisotropic self-similarity
emerges in the flow. The analysis for the Euler equation coincides with that
for the integrable hydrodynamics (\ref{int}).

Decompose the Jacoby matrix $\hat J$ in the bases of eigenvectors of
the direct ($\hat J\psi^{(n)}>=\lambda_n\psi^{(n)}$) and conjugate
($\tilde\psi^{(n)}\hat J=\lambda_n\tilde\psi^{(n)}$) spectral problems:
\begin{equation}
\label{exp}
J_{ik}\equiv\frac{\partial x_k}{\partial a_i}=
\sum_{n=1}^3\lambda_n\psi^{(n)}_i\tilde\psi^{(n)}_k.
\end{equation}
The two sets of eigenvectors are mutually orthogonal:
$$
(\tilde\psi^{(n)}\cdot\psi^{(m)})=\delta_{nm}.
$$
In a vicinity of the point of collapse
the eigenvectors can be regarded as approximately constant.

Decompose the vectors $\bf x$ and $\nabla_{\bf a}$ in (\ref{exp})
in the respective bases, denoting their components by $X_n$ and $A_n$:
$$
X_n=({\bf x}\cdot\psi^{(n)}),\quad\frac{\partial}{\partial A_n}=
(\tilde\psi^{(n)}\cdot\nabla_a).
$$
The vector $\Delta{\bf a}$ can be represented in terms of $A_n$ as follows:
$$
\Delta a_{\alpha}=\sum_n\psi_{\alpha}^{(n)} |\tilde\psi^{(n)}|^2 A_n.
$$
As a result, (\ref{exp}) can be expressed as
\begin{eqnarray}
\frac{\partial X_1}{\partial A_1}=\tau+\Gamma_{mn}A_mA_n,\\
\frac{\partial X_2}{\partial A_2}=\lambda_2,\quad
\frac{\partial X_3}{\partial A_3}=\lambda_3\label{system},
\end{eqnarray}
where
$$
\Gamma_{mn}=\gamma_{\alpha\beta}(\lambda_2\lambda_3)^{-1}
\psi_{\alpha}^{(n)}\psi_{\beta}^{(m)}|\tilde\psi^{(n)}|^2|\tilde\psi^{(m)}|^2
$$
and $\tau=\alpha(t_0-t)/(\lambda_2\lambda_3)$ is assumed to be small.
Consequently, size reduction along the directions
$\psi^{(2)}$ and $\psi^{(3)}$ is the same as in the auxiliary
$\bf a$-space, i.e., $\tau^{1/2}$, but in the soft direction,
$\psi^{(1)}$, the spatial scale is $\sim\tau^{3/2}$.
Therefore, in terms of new self-similar variables $\zeta_1=X_1/\tau^{3/2}$,
$\zeta_2=X_2/\tau^{1/2}$ and $\zeta_3=X_3/\tau^{1/2}$,
integration of the system yields for $\zeta_2$ and $\zeta_3$ a linear
dependence on ${\bf\eta}=\Delta{\bf a}/\tau^{1/2}$, and for $\zeta_1$ --
a cubic one:
\begin{eqnarray}
\label{system1}
\zeta_1=(1+\Gamma_{ij}\eta_i\eta_j)\eta_1 +
\frac 1 2\Gamma_{1i}\eta_i\eta_1^2+\frac 13\Gamma_{11}\eta_1^3,
\quad i,j=2,3\\
\zeta_2=\lambda_2\eta_2,\quad\zeta_3=\lambda_3\eta_3\label{system2}.
\end{eqnarray}
Together with (\ref{ome}), relations (\ref{system1}) and (\ref{system2})
implicitly define the dependence of $\ome$ on $\bf r$ and $t$. The presence
of two different self-similarities shows, that the spatial vorticity
distribution becomes strongly flattened in the first direction, and
a pancake-like structure is formed for $t\to t_0$. Due to (\ref{1}) and
the degeneracy of the mapping ($J=0$), vorticity $\ome$ lies in the plane
of the pancake. Near the singularity behavior of $\ome$ is defined by the
following self-similar asymptotics:
\begin{equation}
\label{8}
\ome=\tau^{-1}{\mathbf\Omega}(\zeta_1,\zeta_2,\zeta_3).
\end{equation}
In essence, in the above analysis one is concerned with the behavior
of the mapping near a fold, and thus breaking of vortex lines can be
naturally explained within the classical catastrophe theory
(\cite{arn1}; \cite{arn2}).

\pagebreak
\section{Super-weak collapse}

According to the collapse classification of \cite{ZK86}, breaking of vortex
lines is not a weak collapse but a super-weak one, because already a
contribution from the singularity to the enstrophy $I=\int|\ome|^2d\bf r$
characterizing the energy dissipation rate due to viscosity is small,
$\sim\tau^{1/2}$; a contribution to the total energy is $\sim\tau^{3/2}$.
However, the integral $\int|\nabla\ome|^2d\bf r$ is divergent as
$t\to t_0$. Thus, the breaking solution ${\bf v}={\bf v(r},t)$ cannot be
continued beyond $t=t_0$ in the Sobolev space $H^2({\bf R}^3)$
with the norm $\parallel{\bf f}\parallel_q\equiv
(\sum_{q\le2}\int|\nabla^q{\bf f}|^2d{\bf r})^{1/2}$.
According to the theorem proved by \cite{BKM}, this suffices for
\begin{equation}
\label{cr}
\int_0^{t_0}\sup_{\bf r}|\ome|dt=\infty
\end{equation}
to hold. The condition (\ref{cr}) is necessary and sufficient
for collapse in the Euler equation, and it is satisfied for (\ref{8}).

Another restriction follows from the theorem by \cite{CFM},
stating that there is no collapse for any $t\in[0,t_0]$ if
\begin{equation}
\label{9}
\int_0^{t_0}\sup|\nabla\xib|^2dt<\infty,
\end{equation}
where the supremum is over a region near the maximum of vorticity $|\ome|$.
Occurrence of collapse implies divergence of the integral (\ref{9}) for
$\tau\to 0$. Consequently, $\sup|\nabla\xib|$ has to increase at least like
$\tau^{-1/2}$.
It is evident that, due to solenoidality of $\ome$, either the derivative
$(\xib\cdot\nabla)\xib$ in the direction along the vector $\ome$
in the pancake-like region should have no a singularity at the scales
of the order of $\tau^{1/2}$ or larger, or the singularity should be weaker
than $\tau^{-1/2}$. However, this does not rule out large gradients of
$\xib$ in a region separated in the soft direction from the pancake-like
region, for instance, with the behavior \hbox{$\partial\xib/\partial
X_1\sim\tau^{-\alpha}$} with $1/2\le\alpha<3/2$. This conjecture is plausible,
since transition from the $\bf a$-space to the physical one involves
a significant contraction in the soft direction of the region near the point of
breaking: a sphere of radius $\sim\tau^{1/2}$ is mapped into the pancake-like
region. Thus, a sphere in the \hbox{$\bf r$-space} of radius $\sim\tau^{1/2}$
containing the pancake includes a large preimage of the region outside the
sphere in the $\bf a$-space of radius $\sim\tau^{1/2}$ (the shape of
the preimage is governed by higher order terms in the expansion (\ref{7})~).
Hence in the process of breaking of vortex lines three scales can appear:
$l_1\sim\tau^{3/2}$, $l_{\perp}\sim\tau^{1/2}$ and an intermediate scale
\hbox{$l_{in}\sim\tau^{\alpha}$} with \hbox{$1/2\le\alpha<3/2$} (the presence of which assures
that there are no contradictions with the theorem of \cite{CFM}).

\section{Numerical results}

To verify the hypothesis that formation of singularity in the solutions
to the Euler equation can be due to vortex line breaking, we performed
a numerical experiment for the system of equations (\ref{1}-\ref{4}).
Two features of this system are notable. First, in contrast with
the original Euler equation, possessing an infinite number of integrals of
motion -- the Cauchy invariants, -- the system (\ref{1}-\ref{4}) is partially
integrated and therefore contains the Cauchy invariants explicitly. Hence,
while the invariants are guaranteed to be conserved when (\ref{1}-\ref{4})
is solved numerically, it is necessary to test to which extent they are conserved
in the course of direct numerical integration of the Euler equation (\ref{5}).
Second, in the system (\ref{1}-\ref{4}) integration in time (in (\ref{3})~) is
separated from integration over space (in (\ref{4})~), i.e. from inversion of
the operator curl.

The system (\ref{1}-\ref{4}) is considered under the periodicity boundary
conditions and inversion of the operator curl can be performed by the standard
spectral techniques with the use of Fast Fourier Transform. The main difficulty
in numerical integration of the system stems from the necessity of transition
(both direct and inverse) between the variables $\bf r$ and $\bf a$ at each
time step. It was circumvented by the use of two independent grids in the
$\bf r$-space: a moving one (the $\bf R$-grid), the motion of whose points is governed
by (\ref{3}), and a steady regular one (the $\bf r$-grid), which coincides with
the $\bf a$-grid. The numerical algorithm consists of the following steps:

\noindent
($i$) by integrating (\ref{3}) in time, find new
positions of the $\bf R$-grid points;

\noindent
($ii$) compute new values (\ref{1}) of $\ome$ on the $\bf R$-grid
by finite differences;

\noindent
($iii$) by linear interpolation from the values of vorticity at nearby points
of the $\bf R$-grid, determine $\ome$ on the $\bf r$-grid
(for that, for each point of the regular grid it is necessary to find
a tetrahedron, containing the point, whose vertices are the nearest points
of the $\bf R$-grid);

\noindent
($iv$) solve the problem (\ref{4}) to determine flow velocity $\bf v$
on the $\bf r$-grid;

\noindent
($v$) by linear interpolation, determine $\bf v$ on the $\bf R$-grid.

\begin{figure}[ht]
\begin{center}
\includegraphics[width=28.5pc,angle=0]{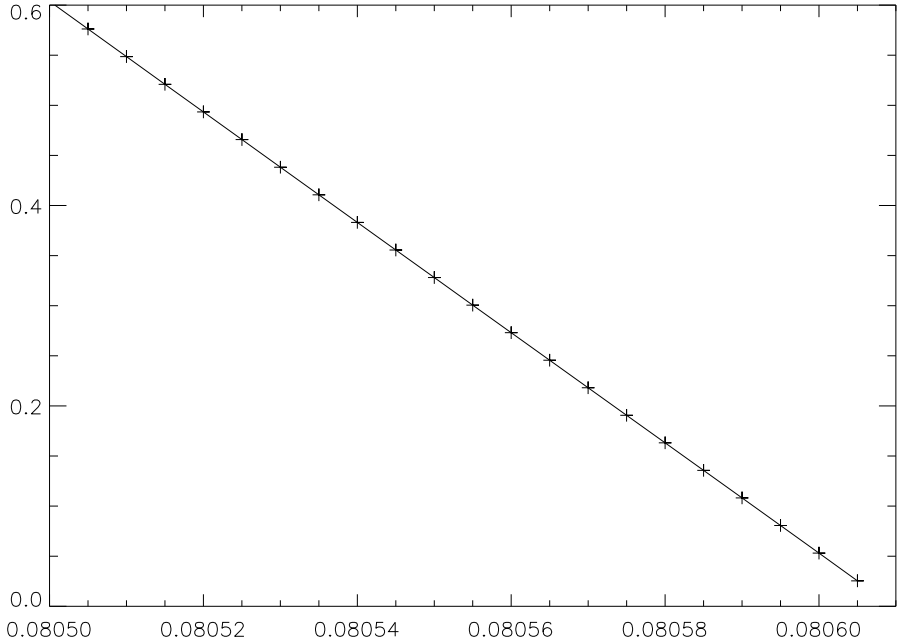}
\end{center}
\caption{The spatial minimum of $|\ome|^{-1}$ (vertical axis)
as a function of time (horizontal axis) at the saturated regime
close to the time of collapse. Pluses show the values obtained in computations.}
\end{figure}

\begin{figure}[ht]
\begin{center}
\includegraphics[width=28.5pc,angle=0]{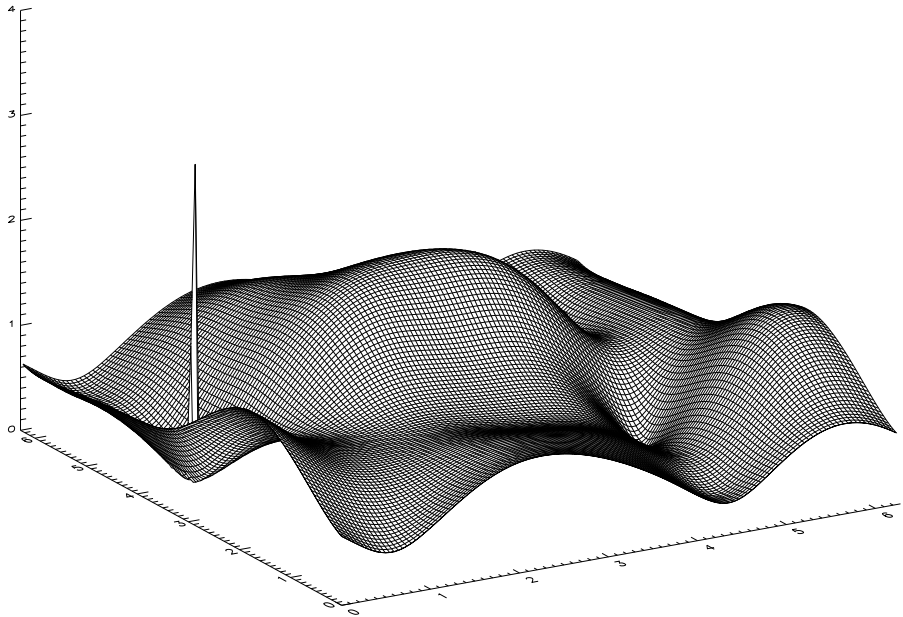}
\end{center}
\caption{Vorticity $|\ome|$ (vertical axis) as a function of the coordinates
$R_1$ and $R_2$ at the plane $R_3=$const, containing the point
of minimum of $J$ at $t=0.08055$ (close to the time of collapse).}
\end{figure}

\begin{figure}[ht]
\begin{center}
\includegraphics[width=28.5pc,angle=0]{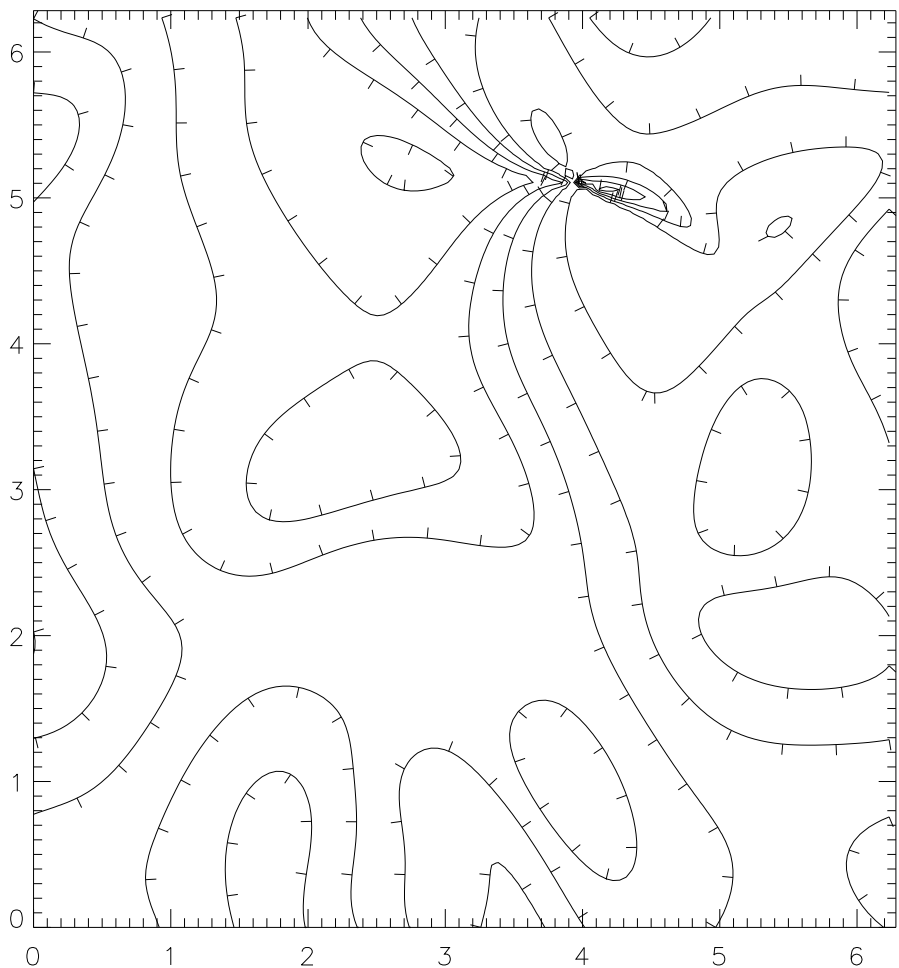}
\end{center}
\caption{Isolines of restriction of the function $R_1({\bf a})$ on
the plane $a_1=$const through the point 
${\bf a}=(7\pi/32; 41\pi/32; 13\pi/8)$, where the collapse occurs.
Small dashes show downhill directons.}
\end{figure}

\begin{figure}[ht]
\begin{center}
\includegraphics[width=28.5pc,angle=0]{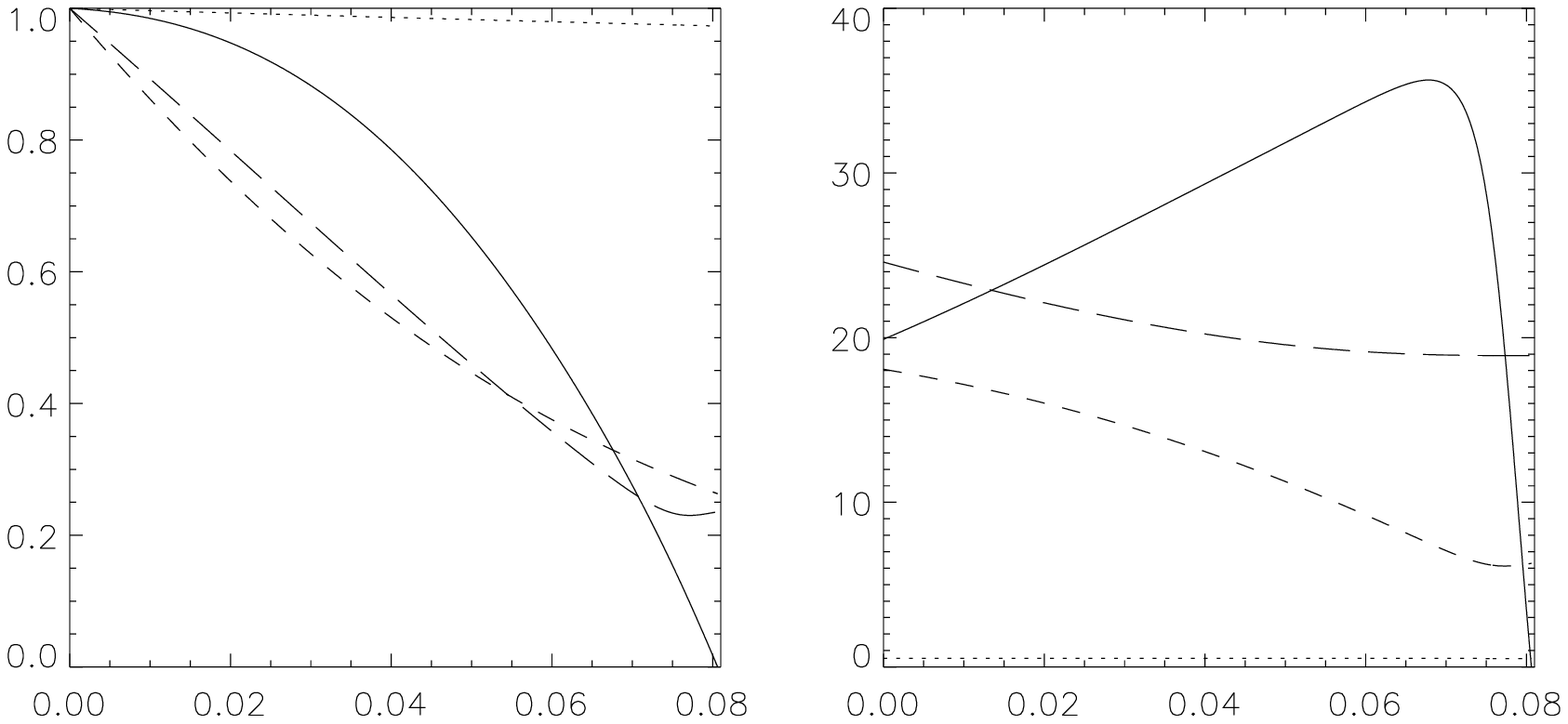}
({\it a})\hspace*{6cm}({\it b})
\end{center}
\caption{Time (horizontal axis) dependencies of four local minima of $J$ (a)
and of $|\ome|^{-1}$ (b). Line styles correspond to the same points
on space. Solid line -- the global minima at the final saturated regime.}
\end{figure}

Computations are performed with the resolution of $128^3$ grid points.
In order to check numerical stability of the algorithm test runs are made
for several initial conditions, which are ABC flows. Any ABC flow is an
eigenfunction of the curl and hence it is a steady solution to the
force-free Euler equation (\ref{5}). They are found to remain steady
in computations with the time step $dt=10^{-3}$ up to $t=4$ with
the relative error of the solution being within the $10^{-14}$ threshold,
and the Jacobian $J$ being reproduced with the $10^{-13}$ accuracy.

An initial vorticity which we consider is a solenoidal field comprised of
random-amplitude Fourier harmonics with an exponentially decaying spectrum;
the decay is by 6 orders of magnitude from the first to the last spherical
shell in the Fourier space, the cut-off being at wavenumber 8.
It satisfies $\ome\ne0$ everywhere in the box of
periodicity (this enables one to perform the projection (\ref{pro}); it is
checked that this condition is not violated at all times during the run).
This field does not possess any symmetry. In the course of numerical
integration we monitor energy conservation: kinetic energy of the flow
remains constant with the accuracy better than 1\%. For such an initial
condition we observe formation of a peak of $|\ome|$ at a {\it single} point.
At this point the Jacobian $J$ and $|\ome|^{-1}$ are minimal over space
at all times close to the time of collapse, and the minimal values decrease in time
to a high precision linearly (Fig.~1). (In this run the time step is
$dt=10^{-4}$ for $t\le0.08$, and $dt=5\cdot10^{-6}$ afterwards.)
In this run the maximum of vorticity increased almost 20
times before integration was terminated. The final width of the peak of
$|\ome|$ is 2-3 times the length of the interval of spatial discretization
(Fig.~2 shows a strong localization of $|\ome|$ at the end of the run).
Figure 3 illustrates concentration of vorticity lines and formation
of a fold near the point of singularity.
Formation of similar peaks of vorticity accompanied by a decrease of $J$ to zero
is also observed in several runs for other initial conditions of vorticity
in the same class.

At different times the global (over space) minima of $J$ and of $|\ome|^{-1}$
are achieved at four different points. Behavior of the Jacobian
in time at one of these points (short-dashed line on Fig.~4) suggests that
the second singularity can also be developing;
it is not traced down to the time of its collapse,
because this is prevented by formation of the first singularity.

The peak of vorticity turns out to be narrow from the moment of its birth.
In order to verify that it is not spurious (i.e. it emerges not due to
a numerical instability of our algorithm) we have reproduced its formation
in computations by a modified algorithm, with different interpolation
techniques employed for linear interpolation at step ($iv$).
These techniques introduce some smoothing intended
to inhibit formation of a spurious singularity.
However, in the new run all numerical data has been reproduced
with the relative precision $10^{-6}$.

\vfill
\pagebreak
To check that the given process can be considered as breaking of
vortex lines we compute time dependencies of the Hessian of the Jacobian
$\partial^2J/\partial a_{\alpha}\partial a_{\beta}$ at the point of the minimum
of $J$. At the final stage of the saturated asymptotic linear behavior of
the minimum we did not find any essential temporal variation of its
eigenvalues. This agrees qualitatively with the expansion (\ref{7}).
Figure 3 illustrates final positions of vortex lines near the point of
collapse. Some anisotropy is observed in the spatial distribution of
$\ome({\bf r},t)$ near the maximum of vorticity. However, due to
an apparent lack of spatial resolution we cannot claim that
two {\it essentially} different scales emerge. The following questions
also remain open: Why is the time of occurrence of collapse
small compared to the turnover time? Why is the peak of vorticity quite
narrow basically from the very moment of its appearance?

The obtained results can be interpreted as the first evidence
of the vortex line breaking; the collapse, which is observed numerically,
is not related to any symmetry of the initial vorticity distribution
and in particular the collapse occurs at a single point.

\begin{acknowledgments}
The authors are grateful to the Observatory of Nice, where this work was
initiated and the paper was completed. Visits of E.K. to the Observatory
of Nice were supported by the Landau-CNRS agreement, and those of O.P.~and
V.Z.~-- by the French Ministry of Education. Participation of E.K.
in the project was also financed by RFBR (grant no.~00-01-00929),
by the Program of Support of the Leading Scientific Schools of Russia
(grant no.~00-15-96007) and by INTAS (grant no. 00-00797).
\end{acknowledgments}

\vfill
\pagebreak

\anxx{Monin, A.S.} \anxx{Yaglom, A.M.} \anxx{L'vov, V.S.}
\anxx{Kolmogorov, A.N.} \anxx{Frisch, U.} \anxx{Kerr, R.M.}
\anxx{Pelz, R.B.} \anxx{Boratav, O.N.} \anxx{Crow, S.C.}
\anxx{Grauer, R.} \anxx{Marliani, C.} \anxx{Germaschewski, K.}
\anxx{Kuznetsov, E.A.} \anxx{Ruban, V.P.} \anxx{Arnold, V.I.}
\anxx{Zakharov, V.E.}
\anxx{Beals, J.T.} \anxx{Kato, T.} \anxx{Majda, A.J.}
\anxx{Constantin, P.} \anxx{Feferman, Ch.} \anxx{Majda, A.J.}

\end{document}